\documentclass{webofc}
\usepackage[varg]{txfonts}   % Web of Conferences font
\usepackage{hyperref}
\usepackage{url}
\hypersetup{colorlinks=true,citecolor=blue,urlcolor=blue,linkcolor=blue}
\usepackage{amsmath}
\usepackage{amssymb}
\usepackage{cleveref}

\newcommand{\RL}{R_\mathrm{L}}

\newcommand{\pt}{p_\mathrm{T}}
\newcommand{\seec}{\Sigma_\mathrm{EEC}}
\newcommand{\sigq}{\Sigma_\mathrm{EEC}^{\mathcal{Q}}}
\newcommand{\ptjet}{p_\mathrm{T, ch\ jet}}

\begin{document}
\title{Charged energy correlators in small systems with ALICE}

\author{\firstname{Tucker} \lastname{Hwang}\inst{1} \fnsep\thanks{\email{tucker_hwang@berkeley.edu}} for the ALICE Collaboration}

\institute{University of California, Berkeley}

\abstract{Energy--energy correlators (EECs), which are energy-weighted cross-sections of particle pairs, offer incisive probes into QCD dynamics, across the full scale of jet evolution, by separating energy scales in the jet fragmentation through the angular distance of the resulting particle pairs. Charged EECs probe the energy flux carried by pairs of the same or opposite electric charges. The interplay between energy distribution and charge conservation enables charged EECs to provide novel constraints on hadronization mechanisms. We present the first measurements of two-point charged energy correlators of inclusive jets in pp collisions at $\sqrt{s} = 5.02$ TeV using the ALICE detector, and compare them with hadronization models to investigate different confinement mechanisms. We also present measurements in p--Pb collisions, examining cold nuclear matter effects on jet evolution.
}
\maketitle
\section{Introduction}
\label{intro}
The energy--energy correlator was originally proposed as an event shape observable in $e^+e^-$ collider experiments for precision measurements of the strong coupling constant \cite{alphas}. More recent measurements at hadron colliders have focused on the collinear limit in jets, where EECs may be able to separate the various energy scales of QCD evolution. The two-point energy correlator (EEC) is defined as a energy-weighted two-particle correlation function:
\begin{equation}
\label{eq:eec}
\Sigma_\mathrm{EEC}(\RL) = \frac{1}{N_\mathrm{jet}}\int \mathrm{d}\RL'\sum_{i,j \in \mathrm{jet}} \frac{p_{\mathrm{T},i}}{\ptjet}\frac{p_{\mathrm{T},j}}{\ptjet}\delta(\RL'-\RL); \quad \RL = \sqrt{\Delta\varphi_{ij}^2 + \Delta\eta_{ij}^2},
\end{equation}
where $i$ and $j$ run over jet constituents, $\ptjet$ is the jet $\pt$, and $N_\mathrm{jet}$ the number of jets. Due to the angular ordering of a jet's parton shower, the EEC probes an energy scale $Q$ proportional to $Q \sim \ptjet\RL$ \cite{eec}. Thus, the EEC has three characteristic regions:
\begin{enumerate}
    \item At small $\RL$, the jet constituents behave like a non-interacting hadron gas. The energies of jet constituents and their angular separation are uncorrelated, and so $\seec \propto \RL$.
    \item At large $\RL$, the EEC probes early-time parton splittings in the shower. Here, perturbative quantum chromodynamics (pQCD) predictions estimate $\seec \propto 1/\RL$ \cite{ceec_th}.
    \item At medium $\RL$ near the peak, the EEC is thought to probe the hadronization transition between hadronic and partonic degrees of freedom.
\end{enumerate}

\section{The charged energy--energy correlator}

The definition in \cref{eq:eec} can be extended to enhance sensitivity to hadronization mechanisms. For example, one can split the jet constituent pairs into different charge categories, termed \textit{charge-selected correlators}. The \textit{unlike-sign correlator} $\Sigma_\mathrm{EEC}^{+-}$ is defined similarly to \cref{eq:eec}, but only pairs $(i, j)$ of unlike-sign charge are considered (and similarly for the \textit{like-sign correlators} $\Sigma_\mathrm{EEC}^{++}$ and $\Sigma_\mathrm{EEC}^{--}$). The \textit{charge-weighted correlator} $\Sigma_\mathrm{EEC}^{\mathcal{Q}}$ is defined as
\begin{equation}
\label{eq:ceec-wt}
\Sigma_\mathrm{EEC}^{\mathcal{Q}}(\RL) = \frac{1}{N_\mathrm{jet}}\int \mathrm{d}\RL'\sum_{i,j \in \mathrm{jet}} q_iq_j\frac{p_{\mathrm{T},i}}{\ptjet}\frac{p_{\mathrm{T},j}}{\ptjet}\delta(\RL'-\RL),
\end{equation}
where the product of the charges are included in the energy weight. Collectively, the charge-weighted and charge-selected energy correlators are termed \textit{charged EECs}.  Note that the charge-weighted correlator can be rewritten as $\Sigma_\mathrm{EEC}^{\mathcal{Q}} = \Sigma_\mathrm{EEC}^{++} + \Sigma_\mathrm{EEC}^{--} - \Sigma_\mathrm{EEC}^{+-}$. The charge-weighted correlator can be interpreted as the overall balance between like-sign and unlike-sign pairs, with like-sign pairs contributing positive weights and vice versa.\footnote{This formalism breaks when considering doubly charged particles, but their contribution to jets is negligible.}

The inclusion of charge information differentially probes energy and charge flow within the confinement transition between partonic and hadronic phases. Charged EECs can thus constrain phenomenological models of hadronization. In addition, EEC measurements in proton--lead (p--Pb) collisions have shown modifications relative to the proton--proton (pp) baseline. Charged EECs are a natural extension of the EEC that can probe the underlying source of these differences, and serve as a new baseline for models.

\section{Experimental setup and analysis techniques}

This analysis uses three subdetectors of the ALICE apparatus: the six-layer silicon Inner Tracking System (ITS) and the Time Projection Chamber (TPC) track charged particles, and the V0 scintillator hodoscopes provide the minimum-bias trigger \cite{perf1,perf2}. ALICE is ideal for jet studies at low $\pt$ due to its good tracking efficiency and momentum resolution at low momenta and small material budget.

These results use data collected in pp and p--Pb collisions during LHC Run 2 at $\sqrt{s_\mathrm{NN}}=5.02$ TeV. Anti-$k_\mathrm{T}$ jets \cite{antikt} with $R=0.4$ are constructed from charged particles\footnote{For the purposes of jet finding, all particles are assumed to have the pion mass.} with $\pt>150$ MeV/$c$ and recombined with the $E$-scheme. Only jets with their full area within the TPC fiducial acceptance ($|\eta_\mathrm{jet}| < 0.9 - R = 0.5$) are considered.  Jet constituents with $\pt > 1$ GeV/$c$ are used to construct particle pairs in the correlator to reduce the contribution from the underlying event (UE). In addition, results from p--Pb (and any corresponding pp reference comparisons) include a jet area-based jet $\pt$ correction \cite{rho}, as well as a perpendicular cone subtraction that removes contributions from pairs that include a particle from the UE.

Detector effects are estimated with simulated events (\textsc{Pythia 8} for pp and DPMJET for p--Pb) passed through a \textsc{Geant3} representation of the ALICE detector. Because the single-track angular resolution is much smaller than the $\RL$ bin widths, no angular bin migrations are considered in the corrections \cite{alice-pp}. Thus, a bin-by-bin correction is employed to account for corrections to $\ptjet$ and energy weights.

\section{Results and discussion}
\begin{figure}[!htb]
\centering
\includegraphics[width=0.75\textwidth]{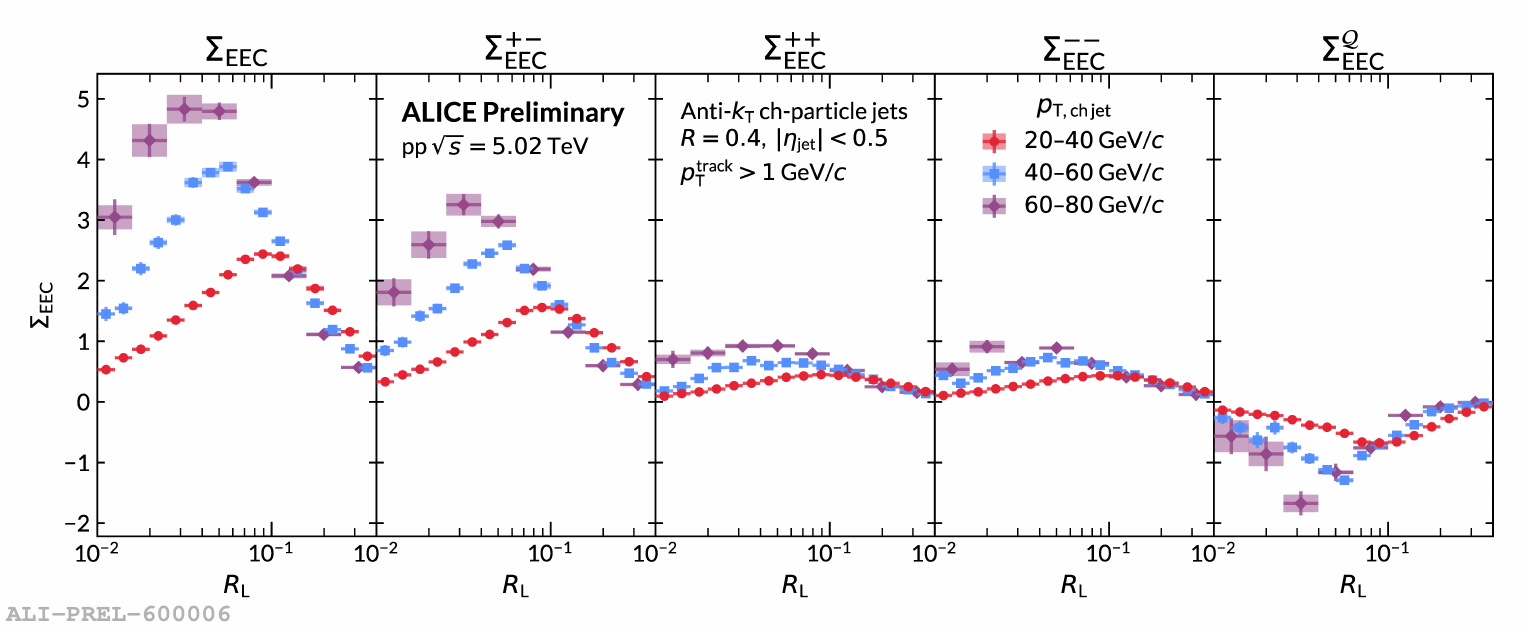}
\includegraphics[width=0.75\textwidth]{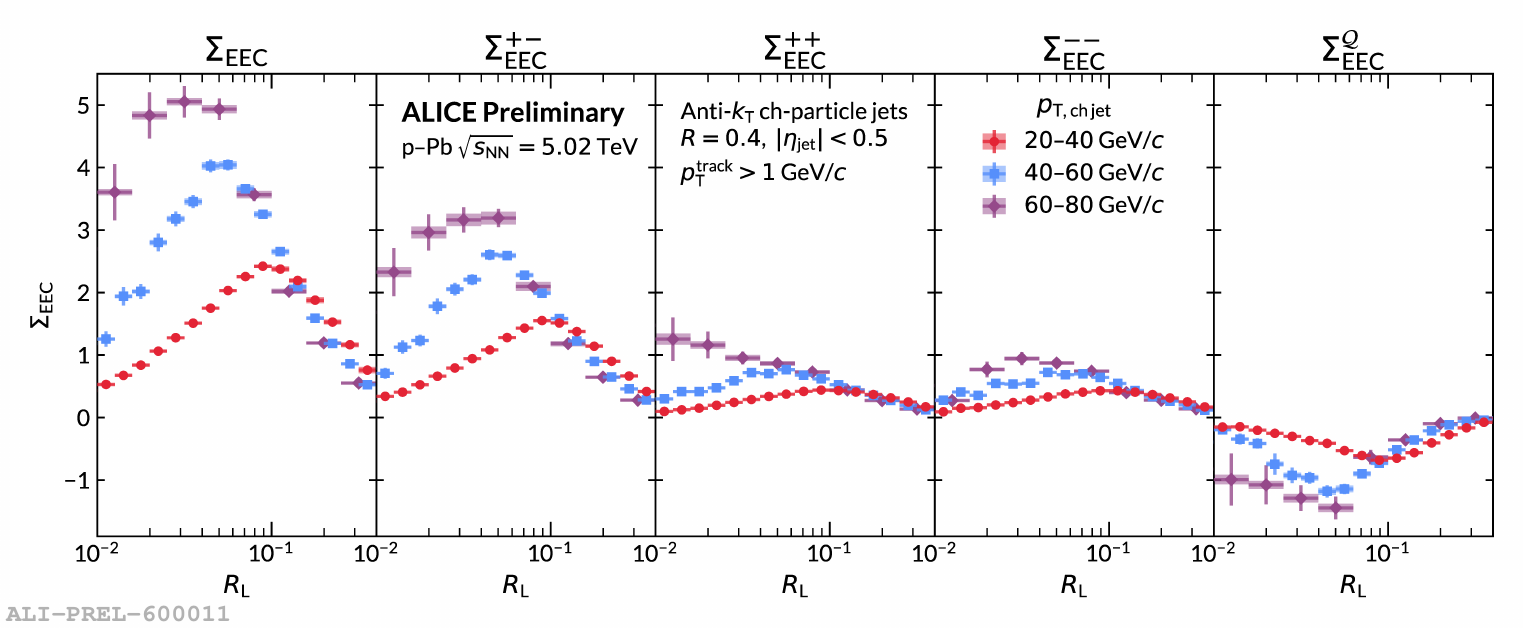}
\caption{The charge-inclusive and charged EECs in pp (upper) and p--Pb (lower) collisions.}
\label{fig:eec-pp}
\end{figure}%
\Cref{fig:eec-pp} shows the charged and inclusive EECs as a function of $\RL$ for pp and p--Pb collisions. Consistent with a universal hadronization energy scale, the angular scale corresponding to the peak of the EEC decreases as the jet $\pt$ increases. A larger contribution from unlike-sign pairs to the charge-inclusive EEC than like-sign pairs is also observed in the relative magnitudes. The symmetry between like-sign correlators indicates that jets fragment equally into positive and negative particles---an expected outcome given QCD isospin symmetry and that gluon-initiated jets dominate inclusive jet samples at LHC energies.%

\begin{figure}[!htb]
  \begin{minipage}[c]{0.75\textwidth}
    \includegraphics[width=\textwidth]{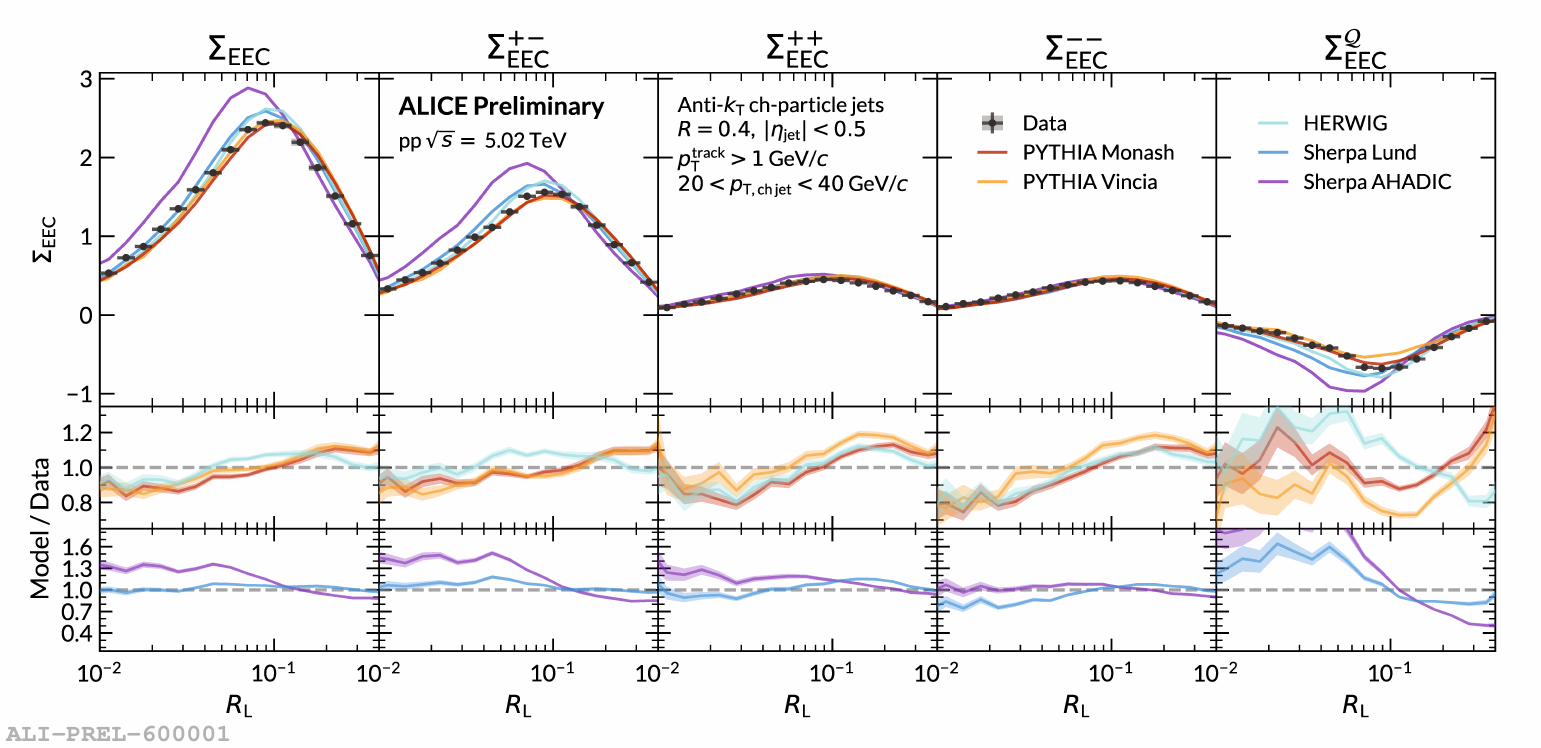}
  \end{minipage}\hfill
  \begin{minipage}[c]{0.24\textwidth}
    \caption{Comparison of charged EECs in pp with five different Monte Carlo models.}
    \label{fig:eec-mc}
  \end{minipage}
\end{figure}%
\Cref{fig:eec-mc} showcases the charged EECs' sensitivity to jet evolution mechanisms. The difference between the string-breaking \textsc{Pythia} and the cluster-hadronizing \textsc{Herwig} in the charge-inclusive EEC is shown to originate from the unlike-sign correlator, rather than the like-sign correlators. On the other hand, different parton shower implementations (\textsc{Pythia} Monash and Vincia) show variations only in like-sign correlations. A direct comparison between string breaking and cluster fragmentation with Sherpa Lund and AHADIC indicates strong sensitivity in the hadronic and confinement regions at small $\RL$.  $\sigq$ exhibits the greatest sensitivity to jet evolution mechanisms, cleanly separating all five models. Ultimately, no models coherently reproduce all the data, and large discrepancies remain.

\begin{figure}[!htb]
  \begin{minipage}[c]{0.5\textwidth}
    \includegraphics[width=0.8\textwidth]{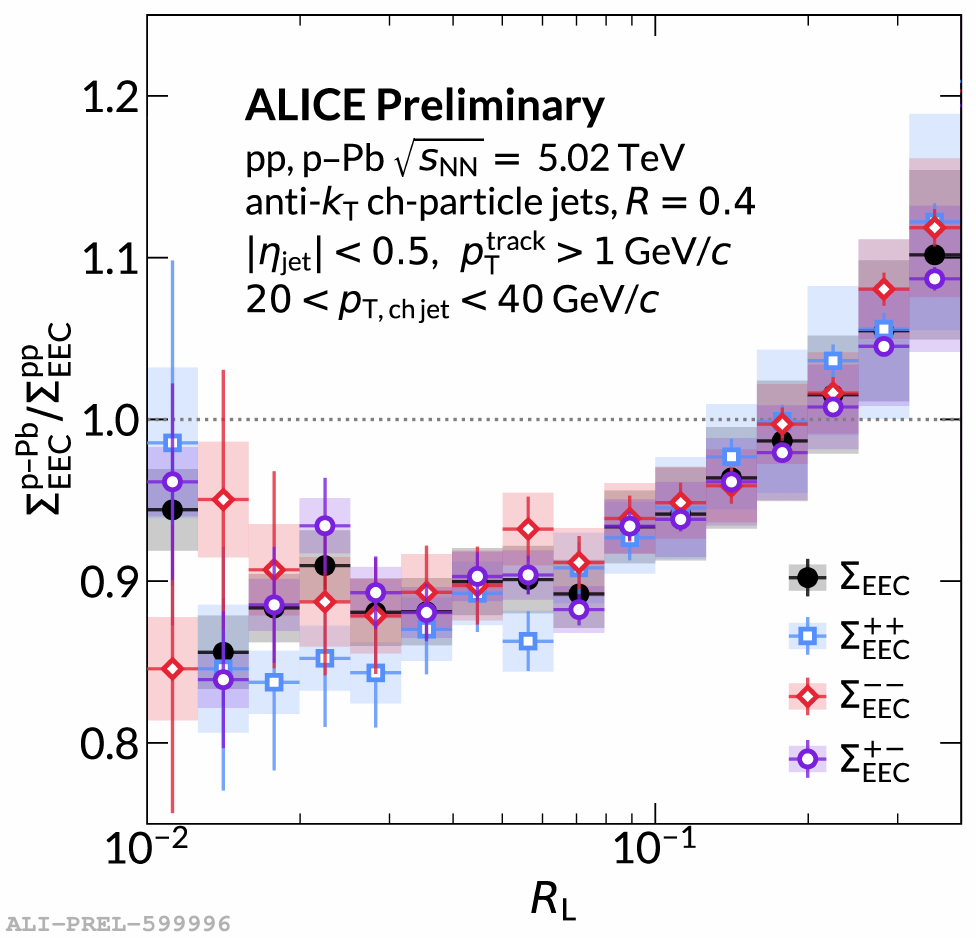}
  \end{minipage}\hfill
  \begin{minipage}[c]{0.5\textwidth}
    \caption{The ratio of charged EECs in p--Pb to pp for inclusive jets with $20 < \ptjet < 40$ GeV/$c$.}
    \label{fig:eec-ratio}
  \end{minipage}
\end{figure}

The charged EECs can also investigate sources of modification in different collision systems. \Cref{fig:eec-ratio} shows that the modification of the EEC in p--Pb collisions relative to pp is identical to that of charged EECs, indicating that the modification is independent of charge.

\section{Conclusions and future prospects}

Charged EECs feature sensitivity to different hadronization and shower mechanisms that are more differentially probed relative to the charge-inclusive EEC. The excellent tracking performance of the ALICE detector at low $\pt$ allows for precise investigation of low energy jets, which have highlighted major discrepancies in how Monte Carlo event generators describe the flow of energy and charge quantum numbers in this jet substructure observable. The heightened sensitivity of charged EECs motivates further studies, in large collision systems (e.g. Pb--Pb) and a wider range of jet samples, such as a quark-enriched jet sample identified via jets recoiling from a photon.

% For tables use syntax in table~\ref{tab-1}.
% \begin{table}
% \centering
% \caption{Please write your table caption here}
% \label{tab-1}       % Give a unique label
% % For LaTeX tables you can use
% \begin{tabular}{lll}
% \hline
% first & second & third  \\\hline
% number & number & number \\
% number & number & number \\
% number & number & number \\\hline
% \end{tabular}
% % Or use
% \vspace*{5cm}  % with the correct table height
% \end{table}
%
% BibTeX or Biber users please use (the style is already called in the class, ensure that the "woc.bst" style is in your local directory)
\bibliography{refs.bib} % Replace "your_bib_file" with the actual name of your .bib file
% \begin{thebibliography}{}
% Journal Author, Article title. Journal \textbf{Volume}, page numbers (year). \url{https://doi.org/Article-DOI-number}
% Book Author, \textit{Book title} (Publisher, place, year) page numbers
% etc
% \end{thebibliography}

\end{document}